\documentclass[aps,prl,twocolumn,superscriptaddress]{revtex4-2}
\usepackage{graphicx} 
\usepackage{dcolumn} 
\usepackage{bm}
\usepackage{color}
\usepackage{hyperref}
\usepackage{orcidlink}
\begin{document}

\title{Measurement of the $g$ factor of ground-state $^{87}$Sr at the parts-per-million level using co-trapped ultracold atoms}

\author{Premjith Thekkeppatt\,\orcidlink{0000-0002-1884-8398}}
 \thanks{Present address: NNF Quantum Computing Programme, Niels Bohr Institute, University of Copenhagen, Universitetsparken 5, 2100 Copenhagen, Denmark}
 \affiliation{Van der Waals-Zeeman Institute, Institute of Physics, University of Amsterdam, Science Park 904, 1098 XH Amsterdam, The Netherlands}
\author{Digvijay\,\orcidlink{0009-0004-4275-4246}}
 \affiliation{Van der Waals-Zeeman Institute, Institute of Physics, University of Amsterdam, Science Park 904, 1098 XH Amsterdam, The Netherlands}
\author{Alexander Urech\,\orcidlink{0000-0003-1663-4705}}
 \affiliation{Van der Waals-Zeeman Institute, Institute of Physics, University of Amsterdam, Science Park 904, 1098 XH Amsterdam, The Netherlands}
  \affiliation{QuSoft, Science Park 123, 1098XG Amsterdam, The Netherlands}
\author{Florian Schreck\,\orcidlink{0000-0001-8225-8803}}
 \affiliation{Van der Waals-Zeeman Institute, Institute of Physics, University of Amsterdam, Science Park 904, 1098 XH Amsterdam, The Netherlands}
  \affiliation{QuSoft, Science Park 123, 1098XG Amsterdam, The Netherlands}
\author{Klaasjan van Druten\,\orcidlink{0000-0003-3326-9447}}
\email[]{87Srgfactor@strontiumBEC.com}
 \affiliation{Van der Waals-Zeeman Institute, Institute of Physics, University of Amsterdam, Science Park 904, 1098 XH Amsterdam, The Netherlands}
  \affiliation{QuSoft, Science Park 123, 1098XG Amsterdam, The Netherlands}

\date{accepted 8 October 2025; this version compiled: \today}

\begin{abstract}
We demonstrate nuclear magnetic resonance of optically trapped ground-state ultracold $^{87}$Sr atoms. Using a scheme in which a cloud of ultracold $^{87}$Rb is co-trapped nearby, we improve the determination of the nuclear $g$ factor, $g_I$, of atomic $^{87}$Sr by more than two orders of magnitude,  reaching accuracy at the parts-per-million level.
We achieve similar accuracy in the ratio of relevant $g$ factors between Rb and Sr. 
This establishes ultracold $^{87}$Sr as an excellent linear in-vacuum magnetometer. More generally, our work demonstrates how ultracold neutral atoms can be used for the precise determination of highly relevant atomic and nuclear magnetic properties. These results are  important for ongoing efforts toward quantum simulation, quantum computation and optical atomic clocks employing $^{87}$Sr, and other neutral alkaline-earth and alkaline-earth-like atoms.
\end{abstract}

\maketitle
Alkaline-earth and alkaline-earth-like atoms are excellent candidates for quantum simulation of many-body physics \cite{Gorshkov2010}, quantum computing \cite{Barnes_assembly_2022} and atom interferometry \cite{Rudolph2020,Fein2020}. They have been pivotal in realizing state-of-the-art optical atomic clocks \cite{ludlow_optical_2015} and molecular clocks \cite{Leung2023}.  Many of these applications strongly rely on having a singlet electronic ground state and long-lived triplet metastable states.
Among these  
atoms, fermionic $^{87}$Sr is especially suited for optical atomic clocks \cite{Takamoto2005,LeTarget2006,Aeppli2024,Zhang2024}, and also stands out because it has the largest
nuclear spin quantum number, $I=9/2$. The tenfold degeneracy arising from this large nuclear spin 
gives rise to SU(10) spin symmetry. $^{87}$Sr is thus an ideal platform to study SU($N$) spin Hamiltonians  with a great degree of control and isolation---for example implementing Kondo lattice models in an optical lattice \cite{Cazalilla2014}. Single $^{87}$Sr atoms trapped in optical tweezers hold great potential for realizing arrays of higher-dimensional qubits, i.e. qudits, with up to ten states \cite{Omanakuttan2021,Omanakuttan2023,Weggemans2022}. A high degree of control in state preparation and manipulation of the nuclear spin states are among the requirements for realizing these systems. 

To control and manipulate the nuclear spin states in $^{87}$Sr coherently, a key quantity is the splitting of the energy levels in the presence of a
magnetic field, the
Zeeman effect. Alkaline-earth(-like) neutral atoms experience weak splittings and/or shifts in an external magnetic field due to the absence of electronic (orbital and spin) angular momentum in the (singlet) electronic ground state. The splitting is determined by the nuclear $g$ factor, $g_I$.
This factor may be written as  $g_I = \mu_I(1-\sigma_d)/\mu_B I$ \cite{Boyd_nuclear_2007}, where $\mu_B$ is the Bohr magneton, 
$\mu_I$ is the magnetic moment of the bare nucleus, and $\sigma_d$ is the diamagnetic shielding factor due to the surrounding electron cloud.
For neutral $^{87}$Sr atoms, the value of $g_I$ is known experimentally at the $10^{-4}$ level \cite{Olschewski1972,Stone2005}.
\begin{figure}[ht]
\includegraphics[width=0.43\textwidth]{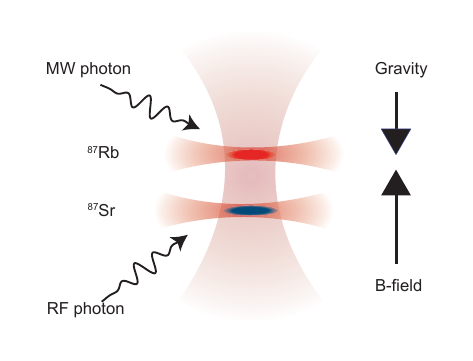} 
\caption{ \label{fig:schematic} 
Schematic of the experimental setup used to measure $g_I$ of $^{87}$Sr in the electronic ground state. Clouds of Rb and Sr atoms are confined in respective crossed optical dipole traps, where the (near-)vertical dipole trap beam is common to both traps. The microwave radiation is emitted using a dedicated microwave antenna, whereas the radio frequency radiation is emitted using coils carrying alternating current.}
\end{figure}
\begin{figure}[ht]
\includegraphics[width=0.46\textwidth]{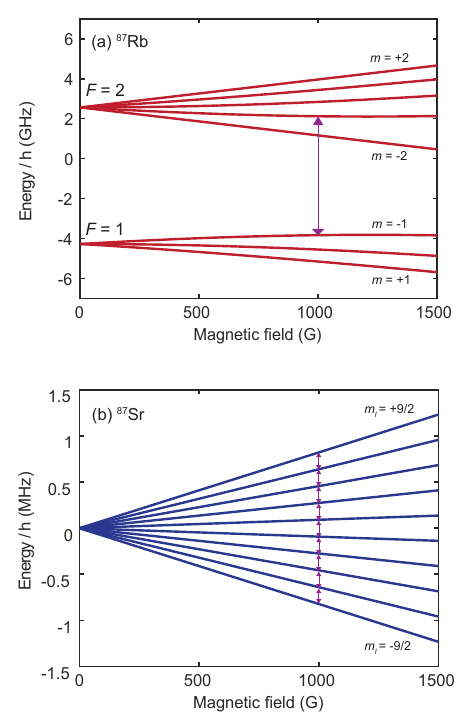}
\caption{ \label{fig:Zeeman} 
(a) Breit-Rabi diagram for $^{87}$Rb in the electronic ground state. The states are labeled by their total magnetic quantum number $m$, and (at low field) by their total angular momentum $F$. The violet arrow marks the hyperfine transition between the two $m=-1$ states used here for (co)magnetometry. (b) Zeeman splitting
of $^{87}$Sr in the electronic ground state. The violet arrows show the transitions between nuclear spin states, labeled by their nuclear magnetic quantum number $m_I$.
Note the differences between (a) and (b), particularly regarding energy scale. These are due to the total electron spin being zero in Sr, while it is nonzero in Rb.
}
\end{figure}

Precision measurement of the $g$ factor of atoms and ions can potentially test quantum electrodynamics (QED) corrections to atomic Hamiltonians involving magnetic interactions. Such measurements are being pursued as a tool for searching for new physics beyond the standard model \cite{Safronova2018,Sturm2013}. By measuring the ratio of the $g$ factor of different isotopes, most of the bound-state QED terms are canceled out, allowing one to probe nuclear effects \cite{Labzowsky1999,Mora2019}. The ratio of the microwave clock transitions of $^{87}$Rb and $^{133}$Cs directly depends on the fine structure constant and the $g$ factor of both atoms and this is used to set limits on the temporal variation of fundamental constants. Improving the accuracy of $g$-factor values enhances the performance of atomic magnetometers, which rely on the measurement of the Zeeman shift to measure the magnetic field. Atomic magnetometers are often used in a comagnetometer configuration together with noble gases or other isotopes, where the samples overlap spatially to gain maximum common-mode sensitivity for the {\em in situ} magnetic field along with the measurement of interest \cite{Wang2020}. Comagnetometers are used for various fundamental physics experiments, such as probing Lorentz and CPT violations, searching for permanent electric dipole moments, and exotic spin-dependent interactions \cite{Kimball2015}.

Here, we report the observation of nuclear magnetic resonance (NMR) of optically trapped ultracold $^{87}$Sr atoms. We have measured NMR transitions at magnetic fields ranging from 100\,G to 1050\,G, resulting in the variation of transition frequencies from 18\,kHz to 194\,kHz. In a comagnetometer-like configuration, we use hyperfine transitions in $^{87}$Rb to determine the magnetic field \textit{in situ}, with mG accuracy. Combining these measurements, we determine the $g$ factor of neutral $^{87}$Sr in its electronic ground state with unprecedented accuracy, at the parts-per-million (ppm) level.

This improves by two orders of magnitude upon the previous state of the art \cite{Banck1973,Sahm1974,Stone2019}, which stood the test of time for over 50 years. Thus, it provides important input for the detailed characterization of the optical clock transition in $^{87}$Sr \cite{Boyd2006,Boyd_nuclear_2007}. 
More generally, this work shows how {\em ultracold neutral atoms} can be used to bring orders of magnitude improvement in the experimental accuracy of highly relevant atomic and nuclear magnetic properties, an area for which typically {\em trapped ions} are used \cite{Schneider2022,Dickopf2024}.

A schematic representation of the experimental arrangement is shown in Fig.~\ref{fig:schematic}. We prepare clouds of ultracold Rb and Sr atoms in vacuum, each in a dedicated optical dipole trap, separated nearly vertically by 0.31(2)\,mm.
For the two traps we use separate horizontal laser beams, combined with a shared (near-)vertical trapping beam (at 1070-nm wavelength) for additional confinement in the  horizontal plane and for mutual alignment of the two respective traps. The horizontal dipole trap beam for Rb is also derived from the 1070-nm laser source, whereas for the Sr trap two crossed horizontal dipole beams with 1064-nm wavelength are used. Both species are loaded from a magneto-optical trap, followed by a short evaporative cooling step.
This results in \(\sim10^{5}\) $^{87}$Rb atoms and \(\sim 4 \times 10^{5} \) $^{87}$Sr atoms in the optical dipole traps, with temperatures of ~1.5\,$\mu$K for $^{87}$Rb and 2.75\,$\mu$K  for $^{87}$Sr. We obtain trap frequencies of 176\,Hz  in the vertical (gravity) direction and 330\,Hz in the horizontal plane for Rb, while for Sr we measure 200\,Hz in the gravity direction and 156\,Hz in the horizontal plane. A more detailed account of the experimental procedure is presented elsewhere \cite{Thek2024b}.  
\begin{figure}[ht]
\includegraphics[width=0.46\textwidth]{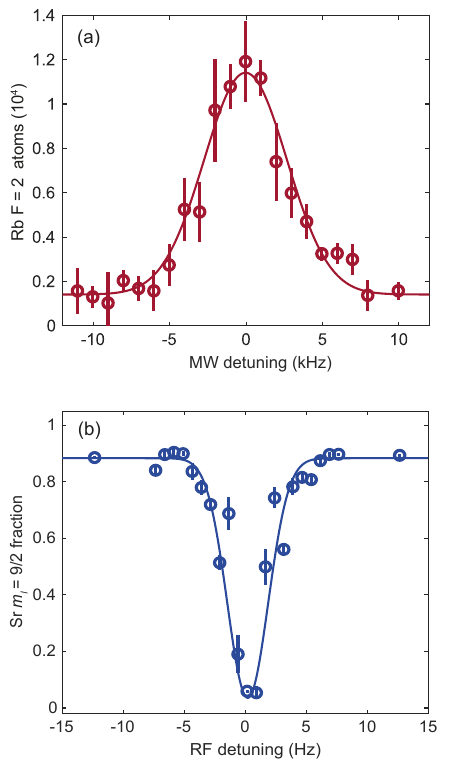}
\caption{\label{fig:spectra}
(a) Spectrum of the $^{87}$Rb hyperfine transition from the lower $m=-1$ state to the upper $m=-1$ state at an inferred magnetic field of 500.544\,G, with Gaussian fit. (b) $^{87}$Sr nuclear magnetic resonance spectrum measured simultaneously in a comagnetometer-like configuration. Solid lines are Gaussian fits to the data.
}
\end{figure}
The Zeeman structures of the electronic ground states of $^{87}$Rb and $^{87}$Sr are illustrated in Fig.\,\ref{fig:Zeeman}. In the experiments, the atoms are prepared in a specific $m$ state. The $^{87}$Rb atoms have $\sim 90\%$ population in the lower $m=-1$ state, while  for $^{87}$Sr, after loading the dipole trap, we optically pump $\sim 95\%$ of the atoms to the $m_I=+9/2$ state. We ramp the magnetic field to the desired value and then perform simultaneous spectroscopy by applying radiofrequency (RF) and microwave (MW) frequency fields to both atomic clouds simultaneously, with pulse durations between 1.5\,s and 3\,s. Because of the vastly different resonant frequencies (see the vertical scales in Fig.\,\ref{fig:Zeeman}), $^{87}$Sr is only sensitive to the RF, while Rb is only sensitive to the MW field. After the spectroscopy pulses, the magnetic field is ramped down on a timescale of 250\,ms; next the atoms are released by switching off the dipole traps. During the subsequent (8~ms) time of flight,
 we employ an optical levitation technique to achieve spin-resolved detection of $^{87}$Sr. Specifically, we selectively levitate the $m_I=+9/2$ state using the scattering force of a laser driving the $\sigma^+$ $^1S_0$\,--\,$^3P_1$ cycling transition \cite{Tey2010}, while the rest of the spin states fall under gravity. 
 A small horizontal offset between the Rb and Sr traps ensures that the Rb does not fall through the center of the optically levitated Sr cloud, avoiding Rb-Sr collisions.
 
 After the time of flight, the Sr clouds are spatially separated enough to distinguish the $m_I=+9/2$ state from the rest of the spin states, and to extract the resulting fraction of Sr atoms in $m_I=+9/2$ by absorption imaging with 461-nm light. For Rb, the atoms transferred to the $F=2$ manifold are imaged using the cooling transition of Rb at 780~nm.
The cycle time of the experiment is $\sim 40$\,s.  

A typical Rb magnetometer spectrum  and the corresponding NMR spectrum  of $^{87}$Sr are shown in Figs.~\ref{fig:spectra}(a) and (b) respectively.
Each data point represents the average of a set of several (typically six) experimental cycles, and error bars represent the standard error of each set. For such spectra, the (near-)resonant RF and MW fields were applied with a square envelope (constant amplitude). We can readily achieve sub-Hz Gaussian linewidths for the Sr NMR transition.

We fit  Gaussian line shapes to the Sr and the Rb spectra. From  the peak position of the fit to the Rb spectrum we extract the magnitude of the local magnetic field at the position of the atoms, while the width of the Rb spectrum is used to determine an upper bound of the residual magnetic field fluctuations over the measurement duration of 1.5 to 2 hours. The data are acquired in a sequential manner, with six scans from low to high frequency. Using the Breit-Rabi formula, the magnetic-field dependence of the transition frequency $f(^{87}$Rb) between the two $m=-1$ hyperfine states used here can be written as 
\begin{equation}
\label{eq:fRb}
    f(^{87}{\rm Rb})=f_0(^{87}{\rm Rb})\sqrt{(1-x)^2+x},
\end{equation}
with $f_0(^{87}{\rm Rb})=6.834\,682\,610\,904\,312\,6(23)$\,Hz, the zero-field (unperturbed) $^{87}$Rb ground-state hyperfine transition frequency \cite{Riehle2018}  and $x=(g_J-g_{I,{\rm Rb}})\mu_BB/hf_0(^{87}{\rm Rb})$, where $h$ is Planck's constant. 
From the accurately known values of the Bohr magneton $\mu_B=eh/4\pi m_e=9.274\,010\,0$657(29)$\times 10^{-28}$\,J/G \cite{CODATA2022} and the relevant fine-structure factor $g_J=2.002\,331\,070(26)$ \cite{Tiedeman1977,CODATA2022} and nuclear factor $g_{I,{\rm Rb}}=-0.000\,995\,1414(10)$ \cite{Arimondo1977}
, the magnetic field strength $B$(Rb) can be readily extracted.

\begin{figure}[ht]
\includegraphics[width=0.45\textwidth]{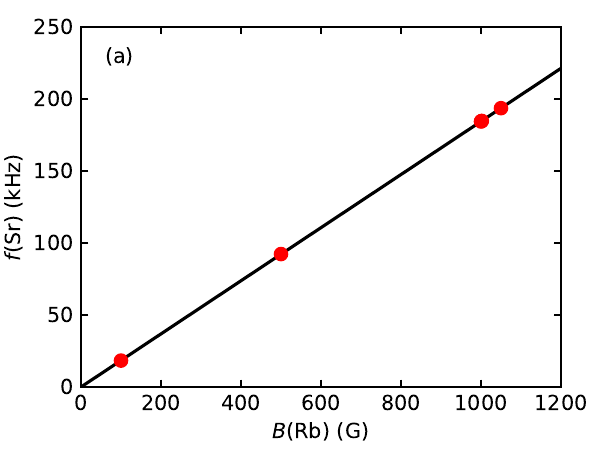}
\includegraphics[width=0.45\textwidth]{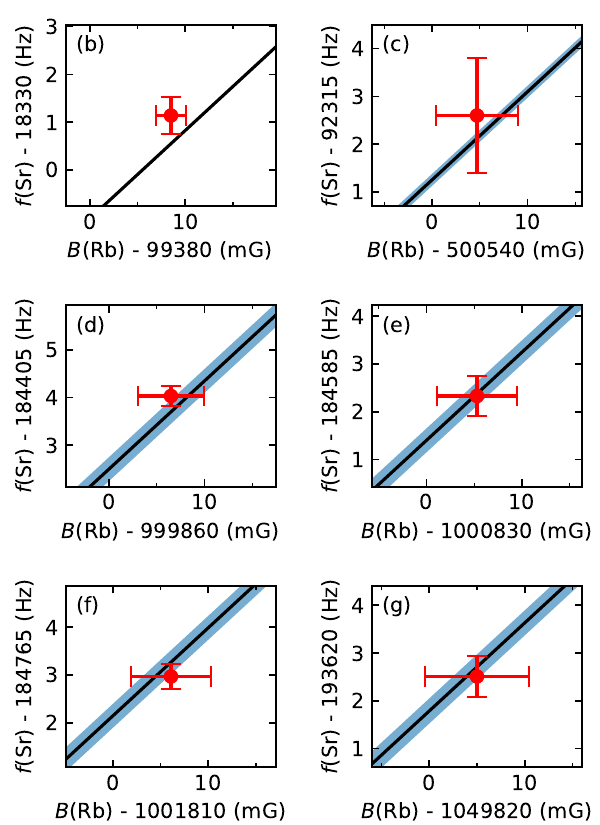}
\caption{\label{fig:summary}
(a) Summary of the $^{87}$Sr $g_I$ data. The measured \(^{87}\text{Sr}\) NMR frequency as a function of the magnitude of the magnetic field $B$ as inferred from the \(^{87}\text{Rb}\) data. The solid line is a linear fit to the data in Table\,\ref{tab:summary}. (b)-(g)  Close-ups on the separate data points including rms widths as error bars, and the linear fit, with the uncertainty range of the fitted slope indicated with a (blue) band.
}
\end{figure}

We perform these measurements at magnetic field settings ranging from 100\,G to 1050\,G to determine the frequency of the $^{87}$Sr NMR transitions as a function of magnetic field.
Table\,\ref{tab:summary} in the End Matter summarizes the resonance frequencies and rms widths obtained from Gaussian fits to the experimental  spectra of both $^{87}$Rb and $^{87}$Sr.  
Taking the magnetic field strengths inferred from the $^{87}$Rb data, and using the NMR transition frequency  $f(^{87}{\rm Sr})=|g_{I,{\rm Sr}}|\mu_B B/h$ between the energy levels
\begin{equation}
\label{eq:fSr}
E_{m_I}({\rm Sr})=- m_I g_{I,{\rm Sr}}\mu_B B,
\end{equation}
the slope of a linear fit gives us the value of $g_I$ of $^{87}$Sr. From a linear fit to these data we obtain a value of 184.43332(25)\,Hz/G, achieving a relative precision of 1.4\,ppm. See also Fig.~\ref{fig:summary} and its discussion in the End Matter.

We now address possible systematic effects in the above determination of the linear Zeeman splitting of ground-state $^{87}$Sr. The main systematic shift arises from the residual magnetic field difference between the locations of the Rb and Sr atoms. Because of the large (near-resonant) s-wave scattering length between $^{87}$Rb and $^{87}$Sr \cite{Barbe2018,Ciamei2018,Thek2024b}, there would be substantial shifts (and three-body losses) if we held the Rb and Sr atoms at the same place at the same time, precluding sufficiently long interrogation times. Instead, as mentioned above, we use a small (sub-mm) displacement of the Rb magnetometer to spatially separate it from the Sr cloud. In this modified comagnetometer arrangement, the frequency measurements are done coincident in time and almost coincident in space. The presence of any stray magnetic field gradient near the atomic cloud thus would lead to a shift in the magnetic field at the position of the Rb cloud compared to the position of the Sr cloud. 

To characterize the strength of the relevant magnetic field difference, we performed magnetometer measurements on Rb at different positions using a movable dipole trap (in the absence of the Sr cloud) and a magnetic field setting of 1001\,G. The position of maximum spatial overlap of the Rb and Sr clouds was determined by scanning the relative position in a separate experiment, maximizing the strong (resonant three-body) losses that occur when Rb and Sr are in contact. We extract a residual gradient of 56(8)\,mG/mm and attribute this to a slight inhomogeneity of the magnetic field from the main coils \cite{Borkowski2023} at the location of the atoms. To account for this inhomogeneity,  we apply a relative correction of $+17(3)\times 10^{-6}$
to $B_{Rb}$ to obtain the magnetic field at the Sr atoms.

Further systematic effects we have considered are differential light shifts and density shifts between the coupled spin states of Rb and similarly for the Sr atoms in our optical dipole traps. These are all found to be negligibly small compared to the one arising from the above magnetic field inhomogeneity. A more detailed discussion can be found in the End Matter, and the results are summarized in Table\,\ref{tab:systematics} there.

Using Eq.~(\ref{eq:fSr}) then leads to 
\begin{equation}
g_{I,{\rm Sr}}=-131.7712(2)_{\rm stat}(3)_{\rm sys} \times 10^{-6}.   
\end{equation}
Expressed as an effective magnetic moment of an isolated neutral $^{87}$Sr atom, this yields $\mu_{\rm eff} = g_{I,{\rm Sr}}\mu_BI=-1.088784(3)\mu_N$, with $\mu_N$ the nuclear magneton.
This value can be compared to the existing reference value for neutral Sr atoms, measured using optical pumping \cite{Olschewski1972,Stone2005}, namely $\mu_{\rm eff}=-1.08859(65)\mu_N$. Our determination thus is consistent with this value, and improves on the relative accuracy of $g_I$ and $\mu_{\rm eff}$ of $^{87}$Sr by more than two orders of magnitude.

Another interesting reference value is the experimental determination of the NMR frequency of $^{87}$Sr ions in aqueous solution, yielding $\mu_{\rm eff}=-1.089 274(7)\mu_N$ \cite{Banck1973,Sahm1974}. For a direct comparison to our data, the difference in diamagnetic shielding $\sigma_d$ between free atoms and ions in solution would have to be known with sufficient accuracy, posing a challenge to the current state of the art of calculating $\sigma_d$. We elaborate on this point in the End Matter. It would also be very interesting to connect our result for $^{87}$Sr to the recent experimental and theoretical determination \cite{Fein2020} of the diamagnetism of the other (naturally occurring, even-numbered) isotopes of Sr, which do not have nuclear spin.

An alternative approach in analyzing our data is to eliminate the magnetic field $B$ and obtain a direct relation between $f(^{87}$Sr) and $f(^{87}$Rb), by substituting $B$ from Eq.\,(\ref{eq:fSr}),  into Eq.\,(\ref{eq:fRb}) (including the above systematic correction to $B$). Fitting the resulting equation to the data leads to a precise determination of the ratio of magnetic moments
\begin{equation}
\label{eq:gIratio}
    \frac{g_{I,{\rm Sr}}}{g_J-g_{I,Rb}}=-65.7762(2)\times 10^{-6},
\end{equation}
with the uncertainty dominated by our present systematic uncertainty in the numerator. The uncertainty in the denominator (i.e. the current relative uncertainty in the literature value of $g_J-g_I$ of ground-state $^{87}$Rb) is at the $10^{-8}$ level, with similar-sized contributions from both $g_J$ and $g_I$ (see values quoted above). We expect that, with improvements of our method and with acquiring further statistics, we should be able to reach the $10^{-8}$ accuracy level in both $g_{I,{\rm Sr}}$ and in the ratio of Eq.~(\ref{eq:gIratio}), i.e., similar to the established accuracy of $g_J$ values in the alkalis \cite{Arimondo1977}.

Future experimental improvements that we envision include bringing the ultracold Rb and Sr closer together ($\approx 0.1$\,mm separation) and, more importantly, closer to the midpoint between the two magnetic coils, suppressing the effect of magnetic field inhomogeneity that is now dominating our systematic uncertainty. Ideally, the two clouds would be at equal distance from the midpoint where the magnetic field magnitude $B$ is at its extremum (minimum), so that the magnetic field experienced by the respective clouds would be equal (up to at least third order in spatial derivatives of $B$).
Acquiring further statistics, in particular at high magnetic fields where the sensitivity is greatest, should allow us to also significantly reduce the statistical uncertainty in $g_{I,{\rm Sr}}$. Some of the other effects that might limit accuracy and precision in next-generation experiments are identified and analyzed in the End Matter.

Our method of NMR on the ten nuclear spin states of optically trapped $^{87}$Sr can be used as a critical tool to prepare coherent superpositions among these states, which otherwise are only accessible via Raman transitions in the optical domain. Further, in the presence of a large tensor light shift, the transition frequency between different nuclear spin states becomes distinct, opening up the possibility of using $^{87}$Sr as a qudecimal with far more quantum control \cite{Omanakuttan2021, Omanakuttan2023}. These transitions can also serve as an essential asset to the toolbox for realizing SU($N$) spin Hamiltonians with alkaline-earth atoms.

In addition to the relevance of magnetic properties of neutral atoms such as $^{87}$Sr for applications in quantum information science and technology, accurate ratios of $g$ factors such as those of Eq.~(\ref{eq:gIratio}) may also offer a different window of opportunity to search for new physics \cite{Safronova2018}--- for instance involving exotic nuclear-spin-dependent couplings \cite{Kimball2015}.
In this context, we point out that Eq.~(\ref{eq:gIratio}) is a ratio of a (predominantly) nuclear property (of $^{87}$Sr) and a (predominantly) electronic one (of $^{87}$Rb). This will be a generally interesting feature for constraining possible variations in (ratios of) fundamental constants of nature, especially if we can improve the accuracy and precision of our data further.

The {\em difference} in linear Zeeman splitting among the $^1S_0$ and $^3P_0$ clock states of $^{87}$Sr has previously been determined  at the 0.4-Hz/G level \cite{Boyd2006,Boyd_nuclear_2007}.
Our determination of the {\em absolute} value of $g_I$ of $^1S_0$  (and hence its linear Zeeman splitting) is accurate at the mHz/G level, and thus can be used  to further constrain the $g$ factor of the $^3P_0$ state. Looking ahead, we expect that we can extend our experimental methods to perform a precision measurement of $g_I$ of $^3P_0$ using NMR of ultracold $^{87}$Sr. We expect to reach similar level of precision as our present determination for $^1S_0$. This could involve using NMR of $^1S_0$ for calibration and accuracy (i.e. eliminating the need for Rb as a co-trapped reference species).  A more precise value of the $^3P_0$ $g$ factor is important for characterizing the effects of Zeeman splittings and shifts at the present and near-future levels of systematic uncertainty for optical atomic clocks based on $^{87}$Sr \cite{Aeppli2024}.

In conclusion, we have measured the $g_I$ factor of neutral $^{87}$Sr using NMR with an unprecedented accuracy at the ppm level using a modified comagnetometer scheme. This reduces the uncertainty in this $g_I$ by more than two orders of magnitude, and poses a challenge to calculations of the diamagnetic shielding factor $\sigma_d$. 
This determination of $g_I$ establishes $^{87}$Sr as an excellent linear in-vacuum magnetometer, and will be  important for applications in quantum simulation, computing, metrology and sensing with fermionic neutral strontium, and in particular for evaluating the magnetic sensitivity of optical lattice clocks based on these atoms. Our work shows how ultracold neutral atoms can be used for precise and accurate determination of highly relevant atomic and nuclear magnetic data. These methods should be readily applicable to other alkaline-earth and alkaline-earth-like atoms.

{\em Acknowledgments}---We thank R.J.C. Spreeuw and the participants of the ICOLS 2025 conference for stimulating discussions. 
This research was funded by NWO Programme 
“Atomic quantum simulators 2.0"  (Project No. 680.92.18.05) and EU project MosaiQC (Project No. 860579). This work was supported by the Dutch National Growth Fund (NGF), as part of the Quantum Delta NL programme. The research of AU is partially funded by NWO  Grant No.\~NGF.1623.23.025 (``Qudits in theory and experiment'').
%
%
\newpage
\begin{center}{\bf End Matter}\end{center}
\begin{table*}[ht]
    \caption{Summary of the experimental results. Nominal magnetic field $B_{\rm nom}$, values of  $^{87}$Rb resonance frequencies $f(^{87}$Rb) and Gaussian root-mean-square (rms) widths $\sigma_f(^{87}$Rb) obtained from fits to the Rb MW spectra,  
    inferred magnetic field values $B$(Rb)
    obtained from these (via the Breit-Rabi formula for Rb),  $^{87}$Sr NMR transition frequencies $f(^{87}$Sr), and corresponding Gaussian rms widths $\sigma_f(^{87}$Sr), and upper bounds $\sigma_B$ to the magnetic field fluctuations as inferred from $\sigma_f({\rm Rb})$ and $\sigma_f({\rm Sr})$, respectively.}
    \label{tab:summary}
    \begin{ruledtabular}
    \begin{tabular}{cccccccc}
    $B_{\rm nom}$ (G) & $f(^{87}$Rb) (kHz)  & $\sigma_f(^{87}$Rb) (kHz) & $B$(Rb) (G) & $\sigma_B({\rm Rb})$ (mG) &  $\sigma_B({\rm Sr})$ (mG)  & $f(^{87}$Sr) (Hz)  & $\sigma_f(^{87}$Sr) (Hz) \\
        \hline
      100  & 6\,699\,692.9  &  2.1 & 99.388\,5 & 1.6 & 2.1 & 18\,331.14  & 0.39 \\
      500  & 6\,252\,223.9  &  3.9 & 500.544\,7 & 4.3 & 7  & 92\,317.6  & 1.2 \\
      1000  & 5\,950\,750.3  &  1.0 & 999.866\,5 & 3.5 & 1.1 & 184\,409.03 & 0.20 \\
      1001 & 5\,950\,470.7  &  1.2 &  1000.835\,3 & 4.2  & 2.2 & 184\,587.33 & 0.41 \\
      1002  & 5\,950\,188.9  &  1.2 &  1001.816\,1 & 4.2 & 1.5 & 184\,767.97 & 0.27 \\
      1050  & 5\,937\,936.2  &  1.2 &  1049.825  & 5.4 & 2.3  & 193\,622.51 & 0.43 
    \end{tabular}
    \end{ruledtabular}
\end{table*}

{\em Overview of $^{87}$Sr NMR data}---Table\,\ref{tab:summary} summarizes the resonance frequencies and rms widths obtained from Gaussian fits to the experimental spectra of both $^{87}$Rb and $^{87}$Sr.
 In Fig.~\ref{fig:summary}, we show an overview of the data along with a linear fit, and close-up views of the data, including widths in the determination of magnetic field and NMR center frequency. Using the transition frequency    $f(^{87}{\rm Sr})=|g_{I,{\rm Sr}}\mu_B B|/h$ between the energy levels, Eq.~(\ref{eq:fSr}), 
the slope of the fit gives us the value of $g_I$ of $^{87}$Sr (before correction for systematic effects).\\
\begin{table*}[tb]
\caption{Summary of  the considered systematic effects for the $^{87}$Rb MW transitions and $^{87}$Sr NMR transitions, and the respective relative effects on the determination of $g_{I,{\rm Sr}}$.}
  \label{tab:systematics}
 \begin{ruledtabular}
 \begin{tabular}{llrr}
 Quantity & Method &
Value & Relative effect on $g_{I,{\rm Sr}}$\\
 \colrule\\[-2mm]
 Magnetic field gradient & Measured &   56(8)\,mG/mm  &  $17(3)\times 10^{-6}$ \\[0.5mm]
 Differential light shifts for $^{87}$Rb & Calculated  &  $\lesssim 100$\,Hz &
  $\lesssim  4\times 10^{-7}$ \\[0.5mm]
 Differential light shifts for $^{87}$Sr & Calculated & $\lesssim 8$\,mHz &  $\lesssim  4\times 10^{-8}$ \\[0.5mm]
 Density shifts for $^{87}$Rb & Calculated & $\lesssim 10$\,Hz & $\lesssim 4\times 10^{-8}$ 
 \end{tabular}
 \end{ruledtabular}
\end{table*}
{\em Systematic effects}---The main systematic effect we found is the difference in magnetic field between Rb and Sr arising from their slightly different positions. This correction and its uncertainty are discussed in the main text. Other systematic effects we have considered are discussed below and the results are summarized in Table\,\ref{tab:systematics}.

The next systematic effect in our experiment comes from the differential light shift  between the coupled spin states of Rb and similarly for the Sr atoms in our optical dipole traps. The nonzero scalar, vector, and tensor differential light shifts were calculated for 1070-nm optical dipole traps, using the methods described in Ref.~\cite{HU18b}. The maximum differential light shift experienced by the $^{87}$Rb hyperfine states is around 100\,Hz. This results in an maximum systematic effect on the determination of $g_{I,{\rm Sr}}$ of $4\times 10^{-7}$.
Our experimental detection threshold for determining this light shift was on the order of the observed linewidth of $\approx 2$\,kHz for the Rb spectra (see Table\,\ref{tab:summary}).

We also evaluated an upper bound to the differential light shifts experienced by different $m_I$ states of $^{87}$Sr in a combined crossed dipole trap of 1070-nm and 1064-nm lasers, and it is estimated to be $\lesssim 8$\,mHz (significantly lower than our experimental linewidths of $\approx 0.5$\,Hz, see Table\,\ref{tab:summary}), leading to a relative effect on $g_{I,{\rm Sr}}$ of $\lesssim 4\times 10^{-8}$. 

A further possible systematic effect is that of density shifts experienced by Rb and Sr atoms. The density shift arises because of the difference in the s-wave scattering lengths  of atoms in various hyperfine states.  Writing the intrastate scattering lengths as $a_{22}$ and $a_{11}$, and the interstate scattering length as $a_{12}$, the density shift $\delta \nu$ in a normal (noncondensed) cloud can be expressed  as 
\begin{equation}
    \delta \nu = \frac{2\hbar}{m}n[a_{\rm 22}-a_{\rm 11}+(2a_{\rm 12}-a_{\rm 11}-a_{\rm 22})f],
\end{equation}
where $n$ is the total density and $f=(n_1-n_2)/n$, with $n_1$ and $n_2$ being individual spin state densities  \cite{harber_effect_2002}.
In the case of $^{87}$Rb, the spin-dependent scattering lengths for different collision channels differ by only a few percent, and are all close to $100a_0$ \cite{harber_effect_2002,Widera_2006,Egorov2013}.  As a result,  we calculate the $^{87}$Rb density shift to be less than 10\,Hz even for the peak density of the Rb cloud, well below our current measurement uncertainty. For $^{87}$Sr we expect the scattering lengths in the electronic ground state  to be independent of the nuclear spin \cite{Gorshkov2010}, and hence a vanishing density shift. 

Methods to reduce the systematic uncertainty due to the magnetic field inhomogeneity have been discussed in the main text. 
There are several possibilities to also reduce the other systematic shifts in our measurements. Differential light shifts can be further suppressed by choosing a Rb dipole trap with magic wavelength for the $^{87}$Rb hyperfine transition. A nearly magic-wavelength optical dipole trap for $^{87}$Rb was demonstrated using a wavelength of 811.5\,nm, where the vector light shifts cancel out the scalar light shifts for certain $m_F$ states \cite{Lundblad2010}.
Density shifts can be avoided if the magnetometer experiment is performed in a three-dimensional optical lattice in the deep Mott insulator regime \cite{Campbell2006}.

We have also considered the possible ac Stark shifts from the RF and microwave radiation on the other (nonresonant) species. The induced shifts are far smaller than our other sources of error. Further, any such shifts can readily be avoided in the future by employing a Ramsey-like spectroscopy scheme, rather than the current Rabi-like excitation.
\\{\em Comparison to reference nuclear data}---As mentioned in the main text, another interesting reference value is the experimental determination through NMR of $\mu_{\rm eff}$ of $^{87}$Sr  ions in aqueous
solution, namely $\mu_{\rm eff}=-1.089 274(7)\mu_N$ \cite{Banck1973,Sahm1974}. For a direct comparison to our data, the difference in diamagnetic shielding $\sigma_d$ between free atoms and ions in solution would have to be known with sufficient accuracy.  To elaborate on this point, in reviews of nuclear data \cite{Fuller1976,Stone2005,Stone2019} it is common to quote the value of magnetic moment of the bare nucleus, $\mu_I$, i.e. after including the (calculated) correction factor for the diamagnetism of the surrounding electron cloud, $\mu_I=\mu_{\rm eff}/(1-\sigma_d)$. However, the uncertainty in the calculated value of $\sigma_d$ is often not clearly stated. For $^{87}$Sr ions in aqueous
solution, a recent calculation \cite{Antusek2013} has an improved value, including uncertainty, of  $\sigma_d=0.00352(11)$, leading to the most recent recommended value of $\mu_I=-1.09316(11)\mu_N$ \cite{Stone2019}, with the uncertainty limited by the calculated value of $\sigma_d$.  

For free neutral atoms, the accepted value for $\sigma_d$ is the so-called LJF (Lin-Johnson-Feiock) correction factor $1/(1-\sigma_d)=1.003950$
\cite{Fuller1976,Stone2005}, although it is not clear what the associated uncertainty is. When combined with our new value of $\mu_{\rm eff}$, this would lead to $\mu_I=-1.093085\mu_N$, i.e. within the uncertainty of the value obtained from aqueous solution. 

This strongly suggests that a more up-to-date and improved calculation of the diamagnetic shielding factor for the free Sr atom, including an estimate of the calculational uncertainty, would allow a much improved accuracy of the magnetic moment of the bare $^{87}$Sr nucleus. When combined with (possibly improved) experimental data for free atoms and in solution, this would constitute a rather stringent test of calculations of diamagnetic shielding factors in these disparate systems.


\begin{thebibliography}{46}%
\makeatletter
\providecommand \@ifxundefined [1]{%
 \@ifx{#1\undefined}
}%
\providecommand \@ifnum [1]{%
 \ifnum #1\expandafter \@firstoftwo
 \else \expandafter \@secondoftwo
 \fi
}%
\providecommand \@ifx [1]{%
 \ifx #1\expandafter \@firstoftwo
 \else \expandafter \@secondoftwo
 \fi
}%
\providecommand \natexlab [1]{#1}%
\providecommand \enquote  [1]{``#1''}%
\providecommand \bibnamefont  [1]{#1}%
\providecommand \bibfnamefont [1]{#1}%
\providecommand \citenamefont [1]{#1}%
\providecommand \href@noop [0]{\@secondoftwo}%
\providecommand \href [0]{\begingroup \@sanitize@url \@href}%
\providecommand \@href[1]{\@@startlink{#1}\@@href}%
\providecommand \@@href[1]{\endgroup#1\@@endlink}%
\providecommand \@sanitize@url [0]{\catcode `\\12\catcode `\$12\catcode
  `\&12\catcode `\#12\catcode `\^12\catcode `\_12\catcode `\%12\relax}%
\providecommand \@@startlink[1]{}%
\providecommand \@@endlink[0]{}%
\providecommand \url  [0]{\begingroup\@sanitize@url \@url }%
\providecommand \@url [1]{\endgroup\@href {#1}{\urlprefix }}%
\providecommand \urlprefix  [0]{URL }%
\providecommand \Eprint [0]{\href }%
\providecommand \doibase [0]{https://doi.org/}%
\providecommand \selectlanguage [0]{\@gobble}%
\providecommand \bibinfo  [0]{\@secondoftwo}%
\providecommand \bibfield  [0]{\@secondoftwo}%
\providecommand \translation [1]{[#1]}%
\providecommand \BibitemOpen [0]{}%
\providecommand \bibitemStop [0]{}%
\providecommand \bibitemNoStop [0]{.\EOS\space}%
\providecommand \EOS [0]{\spacefactor3000\relax}%
\providecommand \BibitemShut  [1]{\csname bibitem#1\endcsname}%
\let\auto@bib@innerbib\@empty
\bibitem [{\citenamefont {Gorshkov}\ \emph {et~al.}(2010)\citenamefont
  {Gorshkov}, \citenamefont {Hermele}, \citenamefont {Gurarie}, \citenamefont
  {Xu}, \citenamefont {Julienne}, \citenamefont {Ye}, \citenamefont {Zoller},
  \citenamefont {Demler}, \citenamefont {Lukin},\ and\ \citenamefont
  {Rey}}]{Gorshkov2010}%
  \BibitemOpen
  \bibfield  {author} {\bibinfo {author} {\bibfnamefont {A.~V.}\ \bibnamefont
  {Gorshkov}}, \bibinfo {author} {\bibfnamefont {M.}~\bibnamefont {Hermele}},
  \bibinfo {author} {\bibfnamefont {V.}~\bibnamefont {Gurarie}}, \bibinfo
  {author} {\bibfnamefont {C.}~\bibnamefont {Xu}}, \bibinfo {author}
  {\bibfnamefont {P.~S.}\ \bibnamefont {Julienne}}, \bibinfo {author}
  {\bibfnamefont {J.}~\bibnamefont {Ye}}, \bibinfo {author} {\bibfnamefont
  {P.}~\bibnamefont {Zoller}}, \bibinfo {author} {\bibfnamefont
  {E.}~\bibnamefont {Demler}}, \bibinfo {author} {\bibfnamefont {M.~D.}\
  \bibnamefont {Lukin}},\ and\ \bibinfo {author} {\bibfnamefont {A.~M.}\
  \bibnamefont {Rey}},\ }\bibfield  {title} {\bibinfo {title} {Two-orbital
  {SU($N$)} magnetism with ultracold alkaline-earth atoms},\ }\href
  {https://doi.org/10.1038/nphys1535} {\bibfield  {journal} {\bibinfo
  {journal} {Nat. Phys.}\ }\textbf {\bibinfo {volume} {6}},\ \bibinfo {pages}
  {289} (\bibinfo {year} {2010})}\BibitemShut {NoStop}%
\bibitem [{\citenamefont {Barnes}\ \emph {et~al.}(2022)\citenamefont {Barnes}
  \emph {et~al.}}]{Barnes_assembly_2022}%
  \BibitemOpen
  \bibfield  {author} {\bibinfo {author} {\bibfnamefont {K.}~\bibnamefont
  {Barnes}} \emph {et~al.},\ }\bibfield  {title} {\bibinfo {title} {Assembly
  and coherent control of a register of nuclear spin qubits},\ }
  \href {https://doi.org/10.1038/s41467-022-29977-z}
  {\bibfield {journal} {\bibinfo  {journal} {Nat. Comm.}\ }\textbf {\bibinfo {volume}
  {13}},\ \bibinfo {pages} {2779}
  (\bibinfo {year} {2022})}\BibitemShut {NoStop}%
\bibitem [{\citenamefont {Rudolph}\ \emph {et~al.}(2020)\citenamefont
  {Rudolph}, \citenamefont {Wilkason}, \citenamefont {Nantel}, \citenamefont
  {Swan}, \citenamefont {Holland}, \citenamefont {Jiang}, \citenamefont
  {Garber}, \citenamefont {Carman},\ and\ \citenamefont {Hogan}}]{Rudolph2020}%
  \BibitemOpen
  \bibfield  {author} {\bibinfo {author} {\bibfnamefont {J.}~\bibnamefont
  {Rudolph}}, \bibinfo {author} {\bibfnamefont {T.}~\bibnamefont {Wilkason}},
  \bibinfo {author} {\bibfnamefont {M.}~\bibnamefont {Nantel}}, \bibinfo
  {author} {\bibfnamefont {H.}~\bibnamefont {Swan}}, \bibinfo {author}
  {\bibfnamefont {C.~M.}\ \bibnamefont {Holland}}, \bibinfo {author}
  {\bibfnamefont {Y.}~\bibnamefont {Jiang}}, \bibinfo {author} {\bibfnamefont
  {B.~E.}\ \bibnamefont {Garber}}, \bibinfo {author} {\bibfnamefont {S.~P.}\
  \bibnamefont {Carman}},\ and\ \bibinfo {author} {\bibfnamefont {J.~M.}\
  \bibnamefont {Hogan}},\ }\bibfield  {title} {\bibinfo {title} {Large momentum
  transfer clock atom interferometry on the 689 nm intercombination line of
  strontium},\ }\href {https://doi.org/10.1103/PhysRevLett.124.083604}
  {\bibfield  {journal} {\bibinfo  {journal} {Phys. Rev. Lett.}\ }\textbf
  {\bibinfo {volume} {124}},\ \bibinfo {pages} {083604} (\bibinfo {year}
  {2020})}\BibitemShut {NoStop}%
\bibitem [{\citenamefont {Fein}\ \emph {et~al.}(2020)\citenamefont {Fein},
  \citenamefont {Shayeghi}, \citenamefont {Mairhofer}, \citenamefont {Kiałka},
  \citenamefont {Rieser}, \citenamefont {Geyer}, \citenamefont {Gerlich},\ and\
  \citenamefont {Arndt}}]{Fein2020}%
  \BibitemOpen
  \bibfield  {author} {\bibinfo {author} {\bibfnamefont {Y.~Y.}\ \bibnamefont
  {Fein}}, \bibinfo {author} {\bibfnamefont {A.}~\bibnamefont {Shayeghi}},
  \bibinfo {author} {\bibfnamefont {L.}~\bibnamefont {Mairhofer}}, \bibinfo
  {author} {\bibfnamefont {F.}~\bibnamefont {Kiałka}}, \bibinfo {author}
  {\bibfnamefont {P.}~\bibnamefont {Rieser}}, \bibinfo {author} {\bibfnamefont
  {P.}~\bibnamefont {Geyer}}, \bibinfo {author} {\bibfnamefont
  {S.}~\bibnamefont {Gerlich}},\ and\ \bibinfo {author} {\bibfnamefont
  {M.}~\bibnamefont {Arndt}},\ }\bibfield  {title} {\bibinfo {title}
  {Quantum-assisted measurement of atomic diamagnetism},\ }\href
  {https://doi.org/10.1103/PhysRevX.10.011014} {\bibfield  {journal} {\bibinfo
  {journal} {Phys. Rev. X}\ }\textbf {\bibinfo {volume} {10}},\ \bibinfo
  {pages} {011014} (\bibinfo {year} {2020})}\BibitemShut {NoStop}%
\bibitem [{\citenamefont {Ludlow}\ \emph {et~al.}(2015)\citenamefont {Ludlow},
  \citenamefont {Boyd}, \citenamefont {Ye}, \citenamefont {Peik},\ and\
  \citenamefont {Schmidt}}]{ludlow_optical_2015}%
  \BibitemOpen
  \bibfield  {author} {\bibinfo {author} {\bibfnamefont {A.~D.}\ \bibnamefont
  {Ludlow}}, \bibinfo {author} {\bibfnamefont {M.~M.}\ \bibnamefont {Boyd}},
  \bibinfo {author} {\bibfnamefont {J.}~\bibnamefont {Ye}}, \bibinfo {author}
  {\bibfnamefont {E.}~\bibnamefont {Peik}},\ and\ \bibinfo {author}
  {\bibfnamefont {P.~O.}\ \bibnamefont {Schmidt}},\ }\bibfield  {title}
  {\bibinfo {title} {Optical atomic clocks},\ }\href
  {https://doi.org/10.1103/RevModPhys.87.637} {\bibfield  {journal} {\bibinfo
  {journal} {Rev. Mod. Phys.}\ }\textbf {\bibinfo {volume} {87}},\ \bibinfo
  {pages} {637} (\bibinfo {year} {2015})}\BibitemShut {NoStop}%
\bibitem [{\citenamefont {Leung}\ \emph {et~al.}(2023)\citenamefont {Leung},
  \citenamefont {Iritani}, \citenamefont {Tiberi}, \citenamefont {Majewska},
  \citenamefont {Borkowski}, \citenamefont {Moszynski},\ and\ \citenamefont
  {Zelevinsky}}]{Leung2023}%
  \BibitemOpen
  \bibfield  {author} {\bibinfo {author} {\bibfnamefont {K.~H.}\ \bibnamefont
  {Leung}}, \bibinfo {author} {\bibfnamefont {B.}~\bibnamefont {Iritani}},
  \bibinfo {author} {\bibfnamefont {E.}~\bibnamefont {Tiberi}}, \bibinfo
  {author} {\bibfnamefont {I.}~\bibnamefont {Majewska}}, \bibinfo {author}
  {\bibfnamefont {M.}~\bibnamefont {Borkowski}}, \bibinfo {author}
  {\bibfnamefont {R.}~\bibnamefont {Moszynski}},\ and\ \bibinfo {author}
  {\bibfnamefont {T.}~\bibnamefont {Zelevinsky}},\ }\bibfield  {title}
  {\bibinfo {title} {Terahertz vibrational molecular clock with systematic
  uncertainty at the ${10}^{-14}$ level},\ }\href
  {https://doi.org/10.1103/PhysRevX.13.011047} {\bibfield  {journal} {\bibinfo
  {journal} {Phys. Rev. X}\ }\textbf {\bibinfo {volume} {13}},\ \bibinfo
  {pages} {011047} (\bibinfo {year} {2023})}\BibitemShut {NoStop}%
\bibitem [{\citenamefont {Takamoto}\ \emph {et~al.}(2005)\citenamefont
  {Takamoto}, \citenamefont {Hong}, \citenamefont {Higashi},\ and\
  \citenamefont {Katori}}]{Takamoto2005}%
  \BibitemOpen
  \bibfield  {author} {\bibinfo {author} {\bibfnamefont {M.}~\bibnamefont
  {Takamoto}}, \bibinfo {author} {\bibfnamefont {F.-L.}\ \bibnamefont {Hong}},
  \bibinfo {author} {\bibfnamefont {R.}~\bibnamefont {Higashi}},\ and\ \bibinfo
  {author} {\bibfnamefont {H.}~\bibnamefont {Katori}},\ }\bibfield  {title}
  {\bibinfo {title} {An optical lattice clock},\ }\href
  {https://doi.org/10.1038/nature03541} {\bibfield  {journal} {\bibinfo
  {journal} {Nature (London)}\ }\textbf {\bibinfo {volume} {435}},\ \bibinfo {pages}
  {321} (\bibinfo {year} {2005})}\BibitemShut {NoStop}%
\bibitem [{\citenamefont {Le~Targat}\ \emph {et~al.}(2006)\citenamefont
  {Le~Targat}, \citenamefont {Baillard}, \citenamefont {Fouch\'e},
  \citenamefont {Brusch}, \citenamefont {Tcherbakoff}, \citenamefont {Rovera},\
  and\ \citenamefont {Lemonde}}]{LeTarget2006}%
  \BibitemOpen
  \bibfield  {author} {\bibinfo {author} {\bibfnamefont {R.}~\bibnamefont
  {Le~Targat}}, \bibinfo {author} {\bibfnamefont {X.}~\bibnamefont {Baillard}},
  \bibinfo {author} {\bibfnamefont {M.}~\bibnamefont {Fouch\'e}}, \bibinfo
  {author} {\bibfnamefont {A.}~\bibnamefont {Brusch}}, \bibinfo {author}
  {\bibfnamefont {O.}~\bibnamefont {Tcherbakoff}}, \bibinfo {author}
  {\bibfnamefont {G.~D.}\ \bibnamefont {Rovera}},\ and\ \bibinfo {author}
  {\bibfnamefont {P.}~\bibnamefont {Lemonde}},\ }\bibfield  {title} {\bibinfo
  {title} {Accurate optical lattice clock with $^{87}\mathrm{Sr}$ atoms},\
  }\href {https://doi.org/10.1103/PhysRevLett.97.130801} {\bibfield  {journal}
  {\bibinfo  {journal} {Phys. Rev. Lett.}\ }\textbf {\bibinfo {volume} {97}},\
  \bibinfo {pages} {130801} (\bibinfo {year} {2006})}\BibitemShut {NoStop}%
\bibitem [{\citenamefont {Aeppli}\ \emph {et~al.}(2024)\citenamefont {Aeppli},
  \citenamefont {Kim}, \citenamefont {Warfield}, \citenamefont {Safronova},\
  and\ \citenamefont {Ye}}]{Aeppli2024}%
  \BibitemOpen
  \bibfield  {author} {\bibinfo {author} {\bibfnamefont {A.}~\bibnamefont
  {Aeppli}}, \bibinfo {author} {\bibfnamefont {K.}~\bibnamefont {Kim}},
  \bibinfo {author} {\bibfnamefont {W.}~\bibnamefont {Warfield}}, \bibinfo
  {author} {\bibfnamefont {M.~S.}\ \bibnamefont {Safronova}},\ and\ \bibinfo
  {author} {\bibfnamefont {J.}~\bibnamefont {Ye}},\ }\bibfield  {title}
  {\bibinfo {title} {Clock with $8\times 10^{-19}$ systematic uncertainty},\
  }\href {https://doi.org/10.1103/PhysRevLett.133.023401} {\bibfield  {journal}
  {\bibinfo  {journal} {Phys. Rev. Lett.}\ }\textbf {\bibinfo {volume} {133}},\
  \bibinfo {pages} {023401} (\bibinfo {year} {2024})}\BibitemShut {NoStop}%
\bibitem [{\citenamefont {Zhang}\ \emph {et~al.}(2024)\citenamefont {Zhang},
  \citenamefont {Ooi}, \citenamefont {Higgins}, \citenamefont {Doyle},
  \citenamefont {von~der Wense}, \citenamefont {Beeks}, \citenamefont
  {Leitner}, \citenamefont {Kazakov}, \citenamefont {Li}, \citenamefont
  {Thirolf}, \citenamefont {Schumm},\ and\ \citenamefont {Ye}}]{Zhang2024}%
  \BibitemOpen
  \bibfield  {author} {\bibinfo {author} {\bibfnamefont {C.}~\bibnamefont
  {Zhang}}, \bibinfo {author} {\bibfnamefont {T.}~\bibnamefont {Ooi}}, \bibinfo
  {author} {\bibfnamefont {J.~S.}\ \bibnamefont {Higgins}}, \bibinfo {author}
  {\bibfnamefont {J.~F.}\ \bibnamefont {Doyle}}, \bibinfo {author}
  {\bibfnamefont {L.}~\bibnamefont {von~der Wense}}, \bibinfo {author}
  {\bibfnamefont {K.}~\bibnamefont {Beeks}}, \bibinfo {author} {\bibfnamefont
  {A.}~\bibnamefont {Leitner}}, \bibinfo {author} {\bibfnamefont {G.~A.}\
  \bibnamefont {Kazakov}}, \bibinfo {author} {\bibfnamefont {P.}~\bibnamefont
  {Li}}, \bibinfo {author} {\bibfnamefont {P.~G.}\ \bibnamefont {Thirolf}},
  \bibinfo {author} {\bibfnamefont {T.}~\bibnamefont {Schumm}},\ and\ \bibinfo
  {author} {\bibfnamefont {J.}~\bibnamefont {Ye}},\ }\bibfield  {title}
  {\bibinfo {title} {Frequency ratio of the $^{229m}${Th} nuclear isomeric
  transition and the $^{87}${Sr} atomic clock},\ }\href
  {https://doi.org/10.1038/s41586-024-07839-6} {\bibfield  {journal} {\bibinfo
  {journal} {Nature (London)}\ }\textbf {\bibinfo {volume} {633}},\ \bibinfo {pages}
  {63} (\bibinfo {year} {2024})}\BibitemShut {NoStop}%
\bibitem [{\citenamefont {Cazalilla}\ and\ \citenamefont
  {Rey}(2014)}]{Cazalilla2014}%
  \BibitemOpen
  \bibfield  {author} {\bibinfo {author} {\bibfnamefont {M.~A.}\ \bibnamefont
  {Cazalilla}}\ and\ \bibinfo {author} {\bibfnamefont {A.~M.}\ \bibnamefont
  {Rey}},\ }\bibfield  {title} {\bibinfo {title} {Ultracold {Fermi} gases with
  emergent {SU($N$)} symmetry},\ }\href
  {https://doi.org/10.1088/0034-4885/77/12/124401} {\bibfield  {journal}
  {\bibinfo  {journal} {Rep. Prog. Phys.}\ }\textbf {\bibinfo {volume} {77}},\
  \bibinfo {pages} {124401} (\bibinfo {year} {2014})}\BibitemShut {NoStop}%
\bibitem [{\citenamefont {Omanakuttan}\ \emph {et~al.}(2021)\citenamefont
  {Omanakuttan}, \citenamefont {Mitra}, \citenamefont {Martin},\ and\
  \citenamefont {Deutsch}}]{Omanakuttan2021}%
  \BibitemOpen
  \bibfield  {author} {\bibinfo {author} {\bibfnamefont {S.}~\bibnamefont
  {Omanakuttan}}, \bibinfo {author} {\bibfnamefont {A.}~\bibnamefont {Mitra}},
  \bibinfo {author} {\bibfnamefont {M.~J.}\ \bibnamefont {Martin}},\ and\
  \bibinfo {author} {\bibfnamefont {I.~H.}\ \bibnamefont {Deutsch}},\
  }\bibfield  {title} {\bibinfo {title} {Quantum optimal control of ten-level
  nuclear spin qudits in $^{87}\text{Sr}$},\ }\href
  {https://doi.org/10.1103/PhysRevA.104.L060401} {\bibfield  {journal}
  {\bibinfo  {journal} {Phys. Rev. A}\ }\textbf {\bibinfo {volume} {104}},\
  \bibinfo {pages} {L060401} (\bibinfo {year} {2021})}\BibitemShut {NoStop}%
\bibitem [{\citenamefont {Omanakuttan}\ \emph {et~al.}(2023)\citenamefont
  {Omanakuttan}, \citenamefont {Mitra}, \citenamefont {Meier}, \citenamefont
  {Martin},\ and\ \citenamefont {Deutsch}}]{Omanakuttan2023}%
  \BibitemOpen
  \bibfield  {author} {\bibinfo {author} {\bibfnamefont {S.}~\bibnamefont
  {Omanakuttan}}, \bibinfo {author} {\bibfnamefont {A.}~\bibnamefont {Mitra}},
  \bibinfo {author} {\bibfnamefont {E.~J.}\ \bibnamefont {Meier}}, \bibinfo
  {author} {\bibfnamefont {M.~J.}\ \bibnamefont {Martin}},\ and\ \bibinfo
  {author} {\bibfnamefont {I.~H.}\ \bibnamefont {Deutsch}},\ }\bibfield
  {title} {\bibinfo {title} {Qudit entanglers using quantum optimal control},\
  }\href {https://doi.org/10.1103/prxquantum.4.040333} {\bibfield  {journal}
  {\bibinfo  {journal} {PRX Quantum}\ }\textbf {\bibinfo {volume} {4}},\
  \bibinfo {pages} {040333} (\bibinfo {year} {2023})}\BibitemShut {NoStop}%
\bibitem [{\citenamefont {Weggemans}\ \emph {et~al.}(2022)\citenamefont
  {Weggemans}, \citenamefont {Urech}, \citenamefont {Rausch}, \citenamefont
  {Spreeuw}, \citenamefont {Boucherie}, \citenamefont {Schreck}, \citenamefont
  {Schoutens}, \citenamefont {Minář},\ and\ \citenamefont
  {Speelman}}]{Weggemans2022}%
  \BibitemOpen
  \bibfield  {author} {\bibinfo {author} {\bibfnamefont {J.~R.}\ \bibnamefont
  {Weggemans}}, \bibinfo {author} {\bibfnamefont {A.}~\bibnamefont {Urech}},
  \bibinfo {author} {\bibfnamefont {A.}~\bibnamefont {Rausch}}, \bibinfo
  {author} {\bibfnamefont {R.}~\bibnamefont {Spreeuw}}, \bibinfo {author}
  {\bibfnamefont {R.}~\bibnamefont {Boucherie}}, \bibinfo {author}
  {\bibfnamefont {F.}~\bibnamefont {Schreck}}, \bibinfo {author} {\bibfnamefont
  {K.}~\bibnamefont {Schoutens}}, \bibinfo {author} {\bibfnamefont
  {J.}~\bibnamefont {Minář}},\ and\ \bibinfo {author} {\bibfnamefont
  {F.}~\bibnamefont {Speelman}},\ }\bibfield  {title} {\bibinfo {title}
  {Solving correlation clustering with {QAOA} and a {Rydberg} qudit system: a
  full-stack approach},\ }\href {https://doi.org/10.22331/q-2022-04-13-687}
  {\bibfield  {journal} {\bibinfo  {journal} {Quantum}\ }\textbf {\bibinfo
  {volume} {6}},\ \bibinfo {pages} {687} (\bibinfo {year} {2022})}\BibitemShut
  {NoStop}%
\bibitem [{\citenamefont {Boyd}\ \emph {et~al.}(2007)\citenamefont {Boyd},
  \citenamefont {Zelevinsky}, \citenamefont {Ludlow}, \citenamefont {Blatt},
  \citenamefont {Zanon-Willette}, \citenamefont {Foreman},\ and\ \citenamefont
  {Ye}}]{Boyd_nuclear_2007}%
  \BibitemOpen
  \bibfield  {author} {\bibinfo {author} {\bibfnamefont {M.~M.}\ \bibnamefont
  {Boyd}}, \bibinfo {author} {\bibfnamefont {T.}~\bibnamefont {Zelevinsky}},
  \bibinfo {author} {\bibfnamefont {A.~D.}\ \bibnamefont {Ludlow}}, \bibinfo
  {author} {\bibfnamefont {S.}~\bibnamefont {Blatt}}, \bibinfo {author}
  {\bibfnamefont {T.}~\bibnamefont {Zanon-Willette}}, \bibinfo {author}
  {\bibfnamefont {S.~M.}\ \bibnamefont {Foreman}},\ and\ \bibinfo {author}
  {\bibfnamefont {J.}~\bibnamefont {Ye}},\ }\bibfield  {title} {\bibinfo
  {title} {Nuclear spin effects in optical lattice clocks},\ }\href
  {https://doi.org/10.1103/PhysRevA.76.022510} {\bibfield  {journal} {\bibinfo
  {journal} {Phys. Rev. A}\ }\textbf {\bibinfo {volume} {76}},\ \bibinfo
  {pages} {022510} (\bibinfo {year} {2007})}\BibitemShut {NoStop}%
\bibitem [{\citenamefont {Olschewski}(1972)}]{Olschewski1972}%
  \BibitemOpen
  \bibfield  {author} {\bibinfo {author} {\bibfnamefont {L.}~\bibnamefont
  {Olschewski}},\ }\bibfield  {title} {\bibinfo {title} {Measurement of nuclear
  magnetic dipole moments in free atoms using optical pumping techniques},\
  }\href {https://doi.org/10.1007/BF01400226} {\bibfield  {journal} {\bibinfo
  {journal} {Z. f. Phys.}\ }\textbf {\bibinfo {volume} {249}},\ \bibinfo
  {pages} {205} (\bibinfo {year} {1972})}\BibitemShut {NoStop}%
\bibitem [{\citenamefont {Stone}(2005)}]{Stone2005}%
  \BibitemOpen
  \bibfield  {author} {\bibinfo {author} {\bibfnamefont {N.~J.}\ \bibnamefont
  {Stone}},\ }\bibfield  {title} {\bibinfo {title} {Table of nuclear magnetic
  dipole and electric quadrupole moments},\ }\href
  {https://doi.org/10.1016/j.adt.2005.04.001} {\bibfield  {journal} {\bibinfo
  {journal} {At. Data Nucl. Data Tables}\ }\textbf {\bibinfo {volume} {90}},\
  \bibinfo {pages} {75} (\bibinfo {year} {2005})}\BibitemShut {NoStop}%
\bibitem [{\citenamefont {Safronova}\ \emph {et~al.}(2018)\citenamefont
  {Safronova}, \citenamefont {Budker}, \citenamefont {Demille}, \citenamefont
  {{Jackson Kimball}}, \citenamefont {Derevianko},\ and\ \citenamefont
  {Clark}}]{Safronova2018}%
  \BibitemOpen
  \bibfield  {author} {\bibinfo {author} {\bibfnamefont {M.~S.}\ \bibnamefont
  {Safronova}}, \bibinfo {author} {\bibfnamefont {D.}~\bibnamefont {Budker}},
  \bibinfo {author} {\bibfnamefont {D.}~\bibnamefont {Demille}}, \bibinfo
  {author} {\bibfnamefont {D.~F.}\ \bibnamefont {{Jackson Kimball}}}, \bibinfo
  {author} {\bibfnamefont {A.}~\bibnamefont {Derevianko}},\ and\ \bibinfo
  {author} {\bibfnamefont {C.~W.}\ \bibnamefont {Clark}},\ }\bibfield  {title}
  {\bibinfo {title} {Search for new physics with atoms and molecules},\ }\href
  {https://doi.org/10.1103/RevModPhys.90.025008} {\bibfield  {journal}
  {\bibinfo  {journal} {Rev. Mod. Phys.}\ }\textbf {\bibinfo {volume} {90}},\
  \bibinfo {pages} {025008} (\bibinfo {year} {2018})}\BibitemShut {NoStop}%
\bibitem [{\citenamefont {Sturm}\ \emph {et~al.}(2013)\citenamefont {Sturm},
  \citenamefont {Wagner}, \citenamefont {Kretzschmar}, \citenamefont {Quint},
  \citenamefont {Werth},\ and\ \citenamefont {Blaum}}]{Sturm2013}%
  \BibitemOpen
  \bibfield  {author} {\bibinfo {author} {\bibfnamefont {S.}~\bibnamefont
  {Sturm}}, \bibinfo {author} {\bibfnamefont {A.}~\bibnamefont {Wagner}},
  \bibinfo {author} {\bibfnamefont {M.}~\bibnamefont {Kretzschmar}}, \bibinfo
  {author} {\bibfnamefont {W.}~\bibnamefont {Quint}}, \bibinfo {author}
  {\bibfnamefont {G.}~\bibnamefont {Werth}},\ and\ \bibinfo {author}
  {\bibfnamefont {K.}~\bibnamefont {Blaum}},\ }\bibfield  {title} {\bibinfo
  {title} {$g$-factor measurement of hydrogenlike $^{28}${Si}$^{13+}$ as a
  challenge to {QED} calculations},\ }\href
  {https://doi.org/10.1103/PhysRevA.87.030501} {\bibfield  {journal} {\bibinfo
  {journal} {Phys. Rev. A}\ }\textbf {\bibinfo {volume} {87}},\ \bibinfo
  {pages} {030501} (\bibinfo {year} {2013})}\BibitemShut {NoStop}%
\bibitem [{\citenamefont {Labzowsky}\ \emph {et~al.}(1999)\citenamefont
  {Labzowsky}, \citenamefont {Goidenko},\ and\ \citenamefont
  {Pyykko}}]{Labzowsky1999}%
  \BibitemOpen
  \bibfield  {author} {\bibinfo {author} {\bibfnamefont {L.}~\bibnamefont
  {Labzowsky}}, \bibinfo {author} {\bibfnamefont {I.}~\bibnamefont
  {Goidenko}},\ and\ \bibinfo {author} {\bibfnamefont {P.}~\bibnamefont
  {Pyykko}},\ }\bibfield  {title} {\bibinfo {title} {Estimates of the
  bound-state {QED} contributions to the $g$-factor of valence $n$s electrons
  in alkali metal atoms},\ }\href {www.elsevier.nlrlocaterphysleta} {\bibfield
  {journal} {\bibinfo  {journal} {Phys. Lett. A}\ }\textbf {\bibinfo {volume}
  {258}},\ \bibinfo {pages} {31} (\bibinfo {year} {1999})}\BibitemShut
  {NoStop}%
\bibitem [{\citenamefont {Mora}\ \emph {et~al.}(2019)\citenamefont {Mora},
  \citenamefont {Cobos}, \citenamefont {Fuentes},\ and\ \citenamefont {{Jackson
  Kimball}}}]{Mora2019}%
  \BibitemOpen
  \bibfield  {author} {\bibinfo {author} {\bibfnamefont {J.}~\bibnamefont
  {Mora}}, \bibinfo {author} {\bibfnamefont {A.}~\bibnamefont {Cobos}},
  \bibinfo {author} {\bibfnamefont {D.}~\bibnamefont {Fuentes}},\ and\ \bibinfo
  {author} {\bibfnamefont {D.~F.}\ \bibnamefont {{Jackson Kimball}}},\
  }\bibfield  {title} {\bibinfo {title} {Measurement of the ratio between
  $g$-factors of the ground states of $^{87}\mathrm{Rb}$ and
  $^{85}\mathrm{Rb}$},\ }\href {https://doi.org/10.1002/andp.201800281}
  {\bibfield  {journal} {\bibinfo  {journal} {Ann.\~Phys. (Berlin)}\ }\textbf
  {\bibinfo {volume} {531}},\ \bibinfo {pages} {201800281} (\bibinfo {year}
  {2019})}\BibitemShut {NoStop}%
\bibitem [{\citenamefont {Wang}\ \emph {et~al.}(2020)\citenamefont {Wang},
  \citenamefont {Peng}, \citenamefont {Zhang}, \citenamefont {Luo},
  \citenamefont {Li}, \citenamefont {Xiong}, \citenamefont {Wang},\ and\
  \citenamefont {Guo}}]{Wang2020}%
  \BibitemOpen
  \bibfield  {author} {\bibinfo {author} {\bibfnamefont {Z.}~\bibnamefont
  {Wang}}, \bibinfo {author} {\bibfnamefont {X.}~\bibnamefont {Peng}}, \bibinfo
  {author} {\bibfnamefont {R.}~\bibnamefont {Zhang}}, \bibinfo {author}
  {\bibfnamefont {H.}~\bibnamefont {Luo}}, \bibinfo {author} {\bibfnamefont
  {J.}~\bibnamefont {Li}}, \bibinfo {author} {\bibfnamefont {Z.}~\bibnamefont
  {Xiong}}, \bibinfo {author} {\bibfnamefont {S.}~\bibnamefont {Wang}},\ and\
  \bibinfo {author} {\bibfnamefont {H.}~\bibnamefont {Guo}},\ }\bibfield
  {title} {\bibinfo {title} {Single-species atomic comagnetometer based on
  $^{87}\mathrm{Rb}$ atoms},\ }\href
  {https://doi.org/10.1103/PhysRevLett.124.193002} {\bibfield  {journal}
  {\bibinfo  {journal} {Phys. Rev. Lett.}\ }\textbf {\bibinfo {volume} {124}},\
  \bibinfo {pages} {193002} (\bibinfo {year} {2020})}\BibitemShut {NoStop}%
\bibitem [{\citenamefont {{Jackson Kimball}}(2015)}]{Kimball2015}%
  \BibitemOpen
  \bibfield  {author} {\bibinfo {author} {\bibfnamefont {D.~F.}\ \bibnamefont
  {{Jackson Kimball}}},\ }\bibfield  {title} {\bibinfo {title} {Nuclear spin
  content and constraints on exotic spin-dependent couplings},\ }\href
  {https://doi.org/10.1088/1367-2630/17/7/073008} {\bibfield  {journal}
  {\bibinfo  {journal} {New J. Phys.}\ }\textbf {\bibinfo {volume} {17}},\
  \bibinfo {pages} {073008} (\bibinfo {year} {2015})}\BibitemShut {NoStop}%
\bibitem [{\citenamefont {Banck}\ and\ \citenamefont
  {Schwenk}(1973)}]{Banck1973}%
  \BibitemOpen
  \bibfield  {author} {\bibinfo {author} {\bibfnamefont {J.}~\bibnamefont
  {Banck}}\ and\ \bibinfo {author} {\bibfnamefont {A.}~\bibnamefont
  {Schwenk}},\ }\bibfield  {title} {\bibinfo {title} {$^{87}${Strontium} {NMR}
  studies},\ }\href {https://doi.org/10.1007/BF01394654} {\bibfield  {journal}
  {\bibinfo  {journal} {Z. Phys.}\ }\textbf {\bibinfo {volume} {265}},\
  \bibinfo {pages} {165} (\bibinfo {year} {1973})}\BibitemShut {NoStop}%
\bibitem [{\citenamefont {Sahm}\ and\ \citenamefont
  {Schwenk}(1974)}]{Sahm1974}%
  \BibitemOpen
  \bibfield  {author} {\bibinfo {author} {\bibfnamefont {W.}~\bibnamefont
  {Sahm}}\ and\ \bibinfo {author} {\bibfnamefont {A.}~\bibnamefont {Schwenk}},\
  }\bibfield  {title} {\bibinfo {title} {Precision measurements of magnetic
  moments of nuclei with weak {NMR} signals},\ }\href
  {https://doi.org/10.1515/zna-1974-1209} {\bibfield  {journal} {\bibinfo
  {journal} {Z. Naturforsch.}\ }\textbf {\bibinfo {volume} {29A}},\ \bibinfo
  {pages} {1763} (\bibinfo {year} {1974})}\BibitemShut {NoStop}%
\bibitem [{\citenamefont {Stone}(2019)}]{Stone2019}%
  \BibitemOpen
  \bibfield  {author} {\bibinfo {author} {\bibfnamefont {N.~J.}\ \bibnamefont
  {Stone}},\ }\href {https://doi.org/10.61092/iaea.yjpc-cns6} {\bibinfo {title}
  {Table of recommended nuclear magnetic dipole moments: Part {I}, long-lived
  states}} (\bibinfo {year} {2019}),\ \Eprint
  {https://arxiv.org/abs/https://www-nds.iaea.org/publications/indc/indc-nds-0794.pdf}
  {https://www-nds.iaea.org/publications/indc/indc-nds-0794.pdf} \BibitemShut
  {NoStop}%
\bibitem [{\citenamefont {Boyd}\ \emph {et~al.}(2006)\citenamefont {Boyd},
  \citenamefont {Zelevinsky}, \citenamefont {Ludlow}, \citenamefont {Foreman},
  \citenamefont {Blatt}, \citenamefont {Ido},\ and\ \citenamefont
  {Ye}}]{Boyd2006}%
  \BibitemOpen
  \bibfield  {author} {\bibinfo {author} {\bibfnamefont {M.~M.}\ \bibnamefont
  {Boyd}}, \bibinfo {author} {\bibfnamefont {T.}~\bibnamefont {Zelevinsky}},
  \bibinfo {author} {\bibfnamefont {A.~D.}\ \bibnamefont {Ludlow}}, \bibinfo
  {author} {\bibfnamefont {S.~M.}\ \bibnamefont {Foreman}}, \bibinfo {author}
  {\bibfnamefont {S.}~\bibnamefont {Blatt}}, \bibinfo {author} {\bibfnamefont
  {T.}~\bibnamefont {Ido}},\ and\ \bibinfo {author} {\bibfnamefont
  {J.}~\bibnamefont {Ye}},\ }\bibfield  {title} {\bibinfo {title} {Optical
  atomic coherence at the 1-second time scale},\ }\href
  {https://doi.org/10.1126/science.1133732} {\bibfield  {journal} {\bibinfo
  {journal} {Science}\ }\textbf {\bibinfo {volume} {314}},\ \bibinfo {pages}
  {1430} (\bibinfo {year} {2006})}
  \BibitemShut
  {NoStop}%
\bibitem [{\citenamefont {Schneider}\ \emph {et~al.}(2022)\citenamefont
  {Schneider}, \citenamefont {Sikora}, \citenamefont {Dickopf}, \citenamefont
  {M{\"u}ller}, \citenamefont {Oreshkina}, \citenamefont {Rischka},
  \citenamefont {Valuev}, \citenamefont {Ulmer}, \citenamefont {Walz},
  \citenamefont {Harman}, \citenamefont {Keitel}, \citenamefont {Mooser},\ and\
  \citenamefont {Blaum}}]{Schneider2022}%
  \BibitemOpen
  \bibfield  {author} {\bibinfo {author} {\bibfnamefont {A.}~\bibnamefont
  {Schneider}}, \bibinfo {author} {\bibfnamefont {B.}~\bibnamefont {Sikora}},
  \bibinfo {author} {\bibfnamefont {S.}~\bibnamefont {Dickopf}}, \bibinfo
  {author} {\bibfnamefont {M.}~\bibnamefont {M{\"u}ller}}, \bibinfo {author}
  {\bibfnamefont {N.~S.}\ \bibnamefont {Oreshkina}}, \bibinfo {author}
  {\bibfnamefont {A.}~\bibnamefont {Rischka}}, \bibinfo {author} {\bibfnamefont
  {I.~A.}\ \bibnamefont {Valuev}}, \bibinfo {author} {\bibfnamefont
  {S.}~\bibnamefont {Ulmer}}, \bibinfo {author} {\bibfnamefont
  {J.}~\bibnamefont {Walz}}, \bibinfo {author} {\bibfnamefont {Z.}~\bibnamefont
  {Harman}}, \bibinfo {author} {\bibfnamefont {C.~H.}\ \bibnamefont {Keitel}},
  \bibinfo {author} {\bibfnamefont {A.}~\bibnamefont {Mooser}},\ and\ \bibinfo
  {author} {\bibfnamefont {K.}~\bibnamefont {Blaum}},\ }\bibfield  {title}
  {\bibinfo {title} {Direct measurement of the $^3${He}$^+$ magnetic moments},\
  }\href {https://doi.org/10.1038/s41586-022-04761-7} {\bibfield  {journal}
  {\bibinfo  {journal} {Nature}\ }\textbf {\bibinfo {volume} {606}},\ \bibinfo
  {pages} {878} (\bibinfo {year} {2022})}\BibitemShut {NoStop}%
\bibitem [{\citenamefont {Dickopf}\ \emph {et~al.}(2024)\citenamefont
  {Dickopf}, \citenamefont {Sikora}, \citenamefont {Kaiser}, \citenamefont
  {M{\"u}ller}, \citenamefont {Ulmer}, \citenamefont {Yerokhin}, \citenamefont
  {Harman}, \citenamefont {Keitel}, \citenamefont {Mooser},\ and\ \citenamefont
  {Blaum}}]{Dickopf2024}%
  \BibitemOpen
  \bibfield  {author} {\bibinfo {author} {\bibfnamefont {S.}~\bibnamefont
  {Dickopf}}, \bibinfo {author} {\bibfnamefont {B.}~\bibnamefont {Sikora}},
  \bibinfo {author} {\bibfnamefont {A.}~\bibnamefont {Kaiser}}, \bibinfo
  {author} {\bibfnamefont {M.}~\bibnamefont {M{\"u}ller}}, \bibinfo {author}
  {\bibfnamefont {S.}~\bibnamefont {Ulmer}}, \bibinfo {author} {\bibfnamefont
  {V.~A.}\ \bibnamefont {Yerokhin}}, \bibinfo {author} {\bibfnamefont
  {Z.}~\bibnamefont {Harman}}, \bibinfo {author} {\bibfnamefont {C.~H.}\
  \bibnamefont {Keitel}}, \bibinfo {author} {\bibfnamefont {A.}~\bibnamefont
  {Mooser}},\ and\ \bibinfo {author} {\bibfnamefont {K.}~\bibnamefont
  {Blaum}},\ }\bibfield  {title} {\bibinfo {title} {Precision spectroscopy on
  $^9${Be} overcomes limitations from nuclear structure},\ }\href
  {https://doi.org/10.1038/s41586-024-07795-1} {\bibfield  {journal} {\bibinfo
  {journal} {Nature}\ }\textbf {\bibinfo {volume} {632}},\ \bibinfo {pages}
  {757} (\bibinfo {year} {2024})}\BibitemShut {NoStop}%
\bibitem [{\citenamefont {Thekkeppatt}(2024)}]{Thek2024b}%
  \BibitemOpen
  \bibfield  {author} {\bibinfo {author} {\bibfnamefont {P.}~\bibnamefont
  {Thekkeppatt}},\ }{\bibinfo {title} {A tale of two species: ultracold
  rubidium and strontium}},\ \href
  {https://hdl.handle.net/11245.1/7befa191-dca7-4eda-819f-1645c147baaa} {Ph.D.
  thesis},\ \bibinfo  {school} {University of Amsterdam} (\bibinfo {year}
  {2024})\BibitemShut {NoStop}%
\bibitem [{\citenamefont {Tey}\ \emph {et~al.}(2010)\citenamefont {Tey},
  \citenamefont {Stellmer}, \citenamefont {Grimm},\ and\ \citenamefont
  {Schreck}}]{Tey2010}%
  \BibitemOpen
  \bibfield  {author} {\bibinfo {author} {\bibfnamefont {M.~K.}\ \bibnamefont
  {Tey}}, \bibinfo {author} {\bibfnamefont {S.}~\bibnamefont {Stellmer}},
  \bibinfo {author} {\bibfnamefont {R.}~\bibnamefont {Grimm}},\ and\ \bibinfo
  {author} {\bibfnamefont {F.}~\bibnamefont {Schreck}},\ }\bibfield  {title}
  {\bibinfo {title} {Double-degenerate {Bose}-{Fermi} mixture of strontium},\
  }\href {https://doi.org/10.1103/PhysRevA.82.011608} {\bibfield  {journal}
  {\bibinfo  {journal} {Phys. Rev. A}\ }\textbf {\bibinfo {volume} {82}},\
  \bibinfo {pages} {011608} (\bibinfo {year} {2010})}\BibitemShut {NoStop}%
\bibitem [{\citenamefont {Riehle}\ \emph {et~al.}(2018)\citenamefont {Riehle},
  \citenamefont {Gill}, \citenamefont {Arias},\ and\ \citenamefont
  {Robertsson}}]{Riehle2018}%
  \BibitemOpen
  \bibfield  {author} {\bibinfo {author} {\bibfnamefont {F.}~\bibnamefont
  {Riehle}}, \bibinfo {author} {\bibfnamefont {P.}~\bibnamefont {Gill}},
  \bibinfo {author} {\bibfnamefont {F.}~\bibnamefont {Arias}},\ and\ \bibinfo
  {author} {\bibfnamefont {L.}~\bibnamefont {Robertsson}},\ }\bibfield  {title}
  {\bibinfo {title} {The {CIPM} list of recommended frequency standard values:
  guidelines and procedures},\ }\href
  {https://doi.org/10.1088/1681-7575/aaa302} {\bibfield  {journal} {\bibinfo
  {journal} {Metrologia}\ }\textbf {\bibinfo {volume} {55}},\ \bibinfo {pages}
  {188} (\bibinfo {year} {2018})}\BibitemShut {NoStop}%
\bibitem [{\citenamefont {Mohr}\ \emph {et~al.}(2025)\citenamefont {Mohr},
  \citenamefont {Newell}, \citenamefont {Taylor},\ and\ \citenamefont
  {Tiesinga}}]{CODATA2022}%
  \BibitemOpen
  \bibfield  {author} {\bibinfo {author} {\bibfnamefont {P.~J.}\ \bibnamefont
  {Mohr}}, \bibinfo {author} {\bibfnamefont {D.~B.}\ \bibnamefont {Newell}},
  \bibinfo {author} {\bibfnamefont {B.~N.}\ \bibnamefont {Taylor}},\ and\
  \bibinfo {author} {\bibfnamefont {E.}~\bibnamefont {Tiesinga}},\ }\bibfield
  {title} {\bibinfo {title} {{CODATA} recommended values of the fundamental
  physical constants: 2022},\ }\href
  {https://doi.org/10.1103/RevModPhys.97.025002} {\bibfield  {journal}
  {\bibinfo  {journal} {Rev. Mod. Phys.}\ }\textbf {\bibinfo {volume} {97}},\
  \bibinfo {pages} {025002} (\bibinfo {year} {2025})}\BibitemShut {NoStop}%
\bibitem [{\citenamefont {Tiedeman}\ and\ \citenamefont
  {Robinson}(1977)}]{Tiedeman1977}%
  \BibitemOpen
  \bibfield  {author} {\bibinfo {author} {\bibfnamefont {J.~S.}\ \bibnamefont
  {Tiedeman}}\ and\ \bibinfo {author} {\bibfnamefont {H.~G.}\ \bibnamefont
  {Robinson}},\ }\bibfield  {title} {\bibinfo {title} {Determination of
  ${{g}_{J}(^{1}H, 1^{2}S_{\frac{1}{2}})}/{{g}_{s}(e)}$: Test of
  mass-independent corrections},\ }\href
  {https://doi.org/10.1103/PhysRevLett.39.602} {\bibfield  {journal} {\bibinfo
  {journal} {Phys. Rev. Lett.}\ }\textbf {\bibinfo {volume} {39}},\ \bibinfo
  {pages} {602} (\bibinfo {year} {1977})}\BibitemShut {NoStop}%
\bibitem [{\citenamefont {Arimondo}\ \emph {et~al.}(1977)\citenamefont
  {Arimondo}, \citenamefont {Inguscio},\ and\ \citenamefont
  {Violino}}]{Arimondo1977}%
  \BibitemOpen
  \bibfield  {author} {\bibinfo {author} {\bibfnamefont {E.}~\bibnamefont
  {Arimondo}}, \bibinfo {author} {\bibfnamefont {M.}~\bibnamefont {Inguscio}},\
  and\ \bibinfo {author} {\bibfnamefont {P.}~\bibnamefont {Violino}},\
  }\bibfield  {title} {\bibinfo {title} {Experimental determinations of the
  hyperfine structure in the alkali atoms},\ }\href
  {https://doi.org/10.1103/RevModPhys.49.31} {\bibfield  {journal} {\bibinfo
  {journal} {Rev. Mod. Phys.}\ }\textbf {\bibinfo {volume} {49}},\ \bibinfo
  {pages} {31} (\bibinfo {year} {1977})}\BibitemShut {NoStop}%
\bibitem [{\citenamefont {Barb{\'e}}\ \emph {et~al.}(2018)\citenamefont
  {Barb{\'e}}, \citenamefont {Ciamei}, \citenamefont {Pasquiou}, \citenamefont
  {Reichsöllner}, \citenamefont {Schreck}, \citenamefont {Żuchowski},\ and\
  \citenamefont {Hutson}}]{Barbe2018}%
  \BibitemOpen
  \bibfield  {author} {\bibinfo {author} {\bibfnamefont {V.}~\bibnamefont
  {Barb{\'e}}}, \bibinfo {author} {\bibfnamefont {A.}~\bibnamefont {Ciamei}},
  \bibinfo {author} {\bibfnamefont {B.}~\bibnamefont {Pasquiou}}, \bibinfo
  {author} {\bibfnamefont {L.}~\bibnamefont {Reichsöllner}}, \bibinfo {author}
  {\bibfnamefont {F.}~\bibnamefont {Schreck}}, \bibinfo {author} {\bibfnamefont
  {P.~S.}\ \bibnamefont {Żuchowski}},\ and\ \bibinfo {author} {\bibfnamefont
  {J.~M.}\ \bibnamefont {Hutson}},\ }\bibfield  {title} {\bibinfo {title}
  {Observation of {Feshbach} resonances between alkali and closed-shell
  atoms},\ }\href {https://doi.org/10.1038/s41567-018-0169-x} {\bibfield
  {journal} {\bibinfo  {journal} {Nat. Phys.}\ }\textbf {\bibinfo {volume}
  {14}},\ \bibinfo {pages} {881} (\bibinfo {year} {2018})}\BibitemShut
  {NoStop}%
\bibitem [{\citenamefont {Ciamei}\ \emph {et~al.}(2018)\citenamefont {Ciamei}
  \emph {et~al.}}]{Ciamei2018}%
  \BibitemOpen
  \bibfield  {author} {\bibinfo {author} {\bibfnamefont {A.}~\bibnamefont
  {Ciamei}} \emph {et~al.},\ }\bibfield  {title} {\bibinfo {title} {The {RbSr}
  {$^2\Sigma^+$} ground state investigated via spectroscopy of hot and
  ultracold molecules},\ }\href {https://doi.org/10.1039/C8CP03919D} {\bibfield
   {journal} {\bibinfo  {journal} {Phys. Chem. Chem. Phys.}\ }\textbf {\bibinfo
  {volume} {20}},\ \bibinfo {pages} {26221} (\bibinfo {year}
  {2018})}\BibitemShut {NoStop}%
\bibitem [{\citenamefont {Borkowski}\ \emph {et~al.}(2023)\citenamefont
  {Borkowski}, \citenamefont {Reichsöllner}, \citenamefont {Thekkeppatt},
  \citenamefont {Barbé}, \citenamefont {van Roon}, \citenamefont {van
  Druten},\ and\ \citenamefont {Schreck}}]{Borkowski2023}%
  \BibitemOpen
  \bibfield  {author} {\bibinfo {author} {\bibfnamefont {M.}~\bibnamefont
  {Borkowski}}, \bibinfo {author} {\bibfnamefont {L.}~\bibnamefont
  {Reichsöllner}}, \bibinfo {author} {\bibfnamefont {P.}~\bibnamefont
  {Thekkeppatt}}, \bibinfo {author} {\bibfnamefont {V.}~\bibnamefont {Barbé}},
  \bibinfo {author} {\bibfnamefont {T.}~\bibnamefont {van Roon}}, \bibinfo
  {author} {\bibfnamefont {K.}~\bibnamefont {van Druten}},\ and\ \bibinfo
  {author} {\bibfnamefont {F.}~\bibnamefont {Schreck}},\ }\bibfield  {title}
  {\bibinfo {title} {Active stabilization of kilogauss magnetic fields to the
  ppm level for magnetoassociation on ultranarrow {Feshbach} resonances},\
  }\href {https://doi.org/10.1063/5.0143825} {\bibfield  {journal} {\bibinfo
  {journal} {Rev. Sci. Instrum.}\ }\textbf {\bibinfo {volume} {94}},\ \bibinfo
  {pages} {073202} (\bibinfo {year} {2023})}\BibitemShut {NoStop}%
\bibitem [{\citenamefont {Hu}\ \emph {et~al.}(2018)\citenamefont {Hu},
  \citenamefont {Freier}, \citenamefont {Sun}, \citenamefont {Leykauf},
  \citenamefont {Schkolnik}, \citenamefont {Yang}, \citenamefont {Krutzik},\
  and\ \citenamefont {Peters}}]{HU18b}%
  \BibitemOpen
  \bibfield  {author} {\bibinfo {author} {\bibfnamefont {Q.-Q.}\ \bibnamefont
  {Hu}}, \bibinfo {author} {\bibfnamefont {C.}~\bibnamefont {Freier}}, \bibinfo
  {author} {\bibfnamefont {Y.}~\bibnamefont {Sun}}, \bibinfo {author}
  {\bibfnamefont {B.}~\bibnamefont {Leykauf}}, \bibinfo {author} {\bibfnamefont
  {V.}~\bibnamefont {Schkolnik}}, \bibinfo {author} {\bibfnamefont
  {J.}~\bibnamefont {Yang}}, \bibinfo {author} {\bibfnamefont {M.}~\bibnamefont
  {Krutzik}},\ and\ \bibinfo {author} {\bibfnamefont {A.}~\bibnamefont
  {Peters}},\ }\bibfield  {title} {\bibinfo {title} {Observation of vector and
  tensor light shifts in $^{87}\mathrm{Rb}$ using near-resonant, stimulated
  {Raman} spectroscopy},\ }\href {https://doi.org/10.1103/PhysRevA.97.013424}
  {\bibfield  {journal} {\bibinfo  {journal} {Phys. Rev. A}\ }\textbf {\bibinfo
  {volume} {97}},\ \bibinfo {pages} {013424} (\bibinfo {year}
  {2018})}\BibitemShut {NoStop}%
\bibitem [{\citenamefont {Harber}\ \emph {et~al.}(2002)\citenamefont {Harber},
  \citenamefont {Lewandowski}, \citenamefont {McGuirk},\ and\ \citenamefont
  {Cornell}}]{harber_effect_2002}%
  \BibitemOpen
  \bibfield  {author} {\bibinfo {author} {\bibfnamefont {D.~M.}\ \bibnamefont
  {Harber}}, \bibinfo {author} {\bibfnamefont {H.~J.}\ \bibnamefont
  {Lewandowski}}, \bibinfo {author} {\bibfnamefont {J.~M.}\ \bibnamefont
  {McGuirk}},\ and\ \bibinfo {author} {\bibfnamefont {E.~A.}\ \bibnamefont
  {Cornell}},\ }\bibfield  {title} {\bibinfo {title} {Effect of cold collisions
  on spin coherence and resonance shifts in a magnetically trapped ultracold
  gas},\ }\href {https://doi.org/10.1103/PhysRevA.66.053616} {\bibfield
  {journal} {\bibinfo  {journal} {Phys. Rev. A}\ }\textbf {\bibinfo {volume}
  {66}},\ \bibinfo {pages} {053616} (\bibinfo {year} {2002})}\BibitemShut
  {NoStop}%
\bibitem [{\citenamefont {Widera}\ \emph {et~al.}(2006)\citenamefont {Widera},
  \citenamefont {Gerbier}, \citenamefont {Fölling}, \citenamefont {Gericke},
  \citenamefont {Mandel},\ and\ \citenamefont {Bloch}}]{Widera_2006}%
  \BibitemOpen
  \bibfield  {author} {\bibinfo {author} {\bibfnamefont {A.}~\bibnamefont
  {Widera}}, \bibinfo {author} {\bibfnamefont {F.}~\bibnamefont {Gerbier}},
  \bibinfo {author} {\bibfnamefont {S.}~\bibnamefont {Fölling}}, \bibinfo
  {author} {\bibfnamefont {T.}~\bibnamefont {Gericke}}, \bibinfo {author}
  {\bibfnamefont {O.}~\bibnamefont {Mandel}},\ and\ \bibinfo {author}
  {\bibfnamefont {I.}~\bibnamefont {Bloch}},\ }\bibfield  {title} {\bibinfo
  {title} {Precision measurement of spin-dependent interaction strengths for
  spin-1 and spin-2 $^{87}${Rb} atoms},\ }\href
  {https://doi.org/10.1088/1367-2630/8/8/152} {\bibfield  {journal} {\bibinfo
  {journal} {New J. Phys.}\ }\textbf {\bibinfo {volume} {8}},\ \bibinfo {pages}
  {152} (\bibinfo {year} {2006})}\BibitemShut {NoStop}%
\bibitem [{\citenamefont {Egorov}\ \emph {et~al.}(2013)\citenamefont {Egorov},
  \citenamefont {Opanchuk}, \citenamefont {Drummond}, \citenamefont {Hall},
  \citenamefont {Hannaford},\ and\ \citenamefont {Sidorov}}]{Egorov2013}%
  \BibitemOpen
  \bibfield  {author} {\bibinfo {author} {\bibfnamefont {M.}~\bibnamefont
  {Egorov}}, \bibinfo {author} {\bibfnamefont {B.}~\bibnamefont {Opanchuk}},
  \bibinfo {author} {\bibfnamefont {P.}~\bibnamefont {Drummond}}, \bibinfo
  {author} {\bibfnamefont {B.~V.}\ \bibnamefont {Hall}}, \bibinfo {author}
  {\bibfnamefont {P.}~\bibnamefont {Hannaford}},\ and\ \bibinfo {author}
  {\bibfnamefont {A.~I.}\ \bibnamefont {Sidorov}},\ }\bibfield  {title}
  {\bibinfo {title} {Measurement of $s$-wave scattering lengths in a
  two-component {Bose}-{Einstein} condensate},\ }\href
  {https://doi.org/10.1103/PhysRevA.87.053614} {\bibfield  {journal} {\bibinfo
  {journal} {Phys. Rev. A}\ }\textbf {\bibinfo {volume} {87}},\ \bibinfo
  {pages} {053614} (\bibinfo {year} {2013})}\BibitemShut {NoStop}%
\bibitem [{\citenamefont {Lundblad}\ \emph {et~al.}(2010)\citenamefont
  {Lundblad}, \citenamefont {Schlosser},\ and\ \citenamefont
  {Porto}}]{Lundblad2010}%
  \BibitemOpen
  \bibfield  {author} {\bibinfo {author} {\bibfnamefont {N.}~\bibnamefont
  {Lundblad}}, \bibinfo {author} {\bibfnamefont {M.}~\bibnamefont
  {Schlosser}},\ and\ \bibinfo {author} {\bibfnamefont {J.~V.}\ \bibnamefont
  {Porto}},\ }\bibfield  {title} {\bibinfo {title} {Experimental observation of
  magic-wavelength behavior of $^{87}\mathrm{Rb}$ atoms in an optical
  lattice},\ }\href {https://doi.org/10.1103/PhysRevA.81.031611} {\bibfield
  {journal} {\bibinfo  {journal} {Phys. Rev. A}\ }\textbf {\bibinfo {volume}
  {81}},\ \bibinfo {pages} {031611} (\bibinfo {year} {2010})}\BibitemShut
  {NoStop}%
\bibitem [{\citenamefont {Campbell}\ \emph {et~al.}(2006)\citenamefont
  {Campbell}, \citenamefont {Mun}, \citenamefont {Boyd}, \citenamefont
  {Medley}, \citenamefont {Leanhardt}, \citenamefont {Marcassa}, \citenamefont
  {Pritchard},\ and\ \citenamefont {Ketterle}}]{Campbell2006}%
  \BibitemOpen
  \bibfield  {author} {\bibinfo {author} {\bibfnamefont {G.~K.}\ \bibnamefont
  {Campbell}}, \bibinfo {author} {\bibfnamefont {J.}~\bibnamefont {Mun}},
  \bibinfo {author} {\bibfnamefont {M.}~\bibnamefont {Boyd}}, \bibinfo {author}
  {\bibfnamefont {P.}~\bibnamefont {Medley}}, \bibinfo {author} {\bibfnamefont
  {A.~E.}\ \bibnamefont {Leanhardt}}, \bibinfo {author} {\bibfnamefont {L.~G.}\
  \bibnamefont {Marcassa}}, \bibinfo {author} {\bibfnamefont {D.~E.}\
  \bibnamefont {Pritchard}},\ and\ \bibinfo {author} {\bibfnamefont
  {W.}~\bibnamefont {Ketterle}},\ }\bibfield  {title} {\bibinfo {title}
  {Imaging the {Mott} insulator shells by using atomic clock shifts},\ }\href
  {https://doi.org/10.1126/science.1130365} {\bibfield  {journal} {\bibinfo
  {journal} {Science}\ }\textbf {\bibinfo {volume} {313}},\ \bibinfo {pages}
  {649} (\bibinfo {year} {2006})}\BibitemShut {NoStop}%
\bibitem [{\citenamefont {Fuller}(1976)}]{Fuller1976}%
  \BibitemOpen
  \bibfield  {author} {\bibinfo {author} {\bibfnamefont {G.~H.}\ \bibnamefont
  {Fuller}},\ }\bibfield  {title} {\bibinfo {title} {Nuclear spins and
  moments},\ }\href@noop {} {\bibfield  {journal} {\bibinfo  {journal} {J.
  Phys. Chem. Ref. Data}\ }\textbf {\bibinfo {volume} {5}},\ \bibinfo {pages}
  {835} (\bibinfo {year} {1976})}\BibitemShut {NoStop}%
\bibitem [{\citenamefont {Antušek}\ \emph {et~al.}(2013)\citenamefont
  {Antušek} \emph {et~al.}}]{Antusek2013}%
  \BibitemOpen
  \bibfield  {author} {\bibinfo {author} {\bibfnamefont {A.}~\bibnamefont
  {Antušek}} \emph {et~al.},\ }\bibfield  {title} {\bibinfo {title} {Ab initio
  study of {NMR} shielding of alkali earth metal ions in water complexes and
  magnetic moments of alkali earth metal nuclei},\ }\href
  {https://doi.org/https://doi.org/10.1016/j.cplett.2013.10.018} {\bibfield
  {journal} {\bibinfo  {journal} {Chem. Phys. Lett.}\ }\textbf {\bibinfo
  {volume} {588}},\ \bibinfo {pages} {57} (\bibinfo {year} {2013})}\BibitemShut
  {NoStop}%
\end{thebibliography}
\end{document}